\documentclass[
	a4paper, 
	10pt, 
	unnumberedsections, 
	twoside, 
]{LTJournalArticle}

\usepackage{newtxtext,newtxmath}
\usepackage{amsmath}	

\runninghead{ConTEST} 

\footertext{} 

\title{\textbf{Consistency Tests for Comparing Astrophysical Models and Observations}} 

\author{
	\href{https://orcid.org/0000-0002-3424-8528}{\includegraphics[scale=0.09]{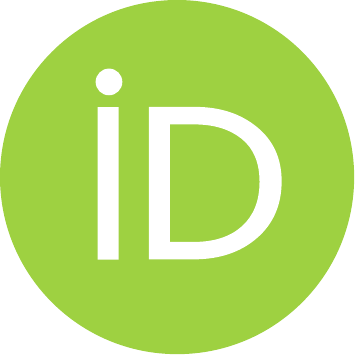}} Fiorenzo Stoppa\textsuperscript{1,2}, 
        \href{https://orcid.org/0000-0003-3873-7983}{\includegraphics[scale=0.09]{orcid.pdf}} Eric Cator\textsuperscript{2},
        \href{https://orcid.org/0000-0002-0752-2974}{\includegraphics[scale=0.09]{orcid.pdf}} Gijs Nelemans\textsuperscript{1,3,4}
}

\date{\footnotesize\textsuperscript{\textbf{1}}Department of Astrophysics/IMAPP, Radboud University, P.O. Box 9010, NL-6500 GL Nijmegen, The Netherlands\\ 
        \textsuperscript{\textbf{2}}Department of Mathematics/IMAPP, Radboud University, P.O. Box 9010, NL-6500 GL Nijmegen, The Netherlands\\
        \textsuperscript{\textbf{3}}Institute of Astronomy, KU Leuven, Celestijnenlaan 200D, B-3001 Leuven, Belgium\\
        \textsuperscript{\textbf{4}}SRON, Netherlands Institute for Space Research, Sorbonnelaan 2, NL-3584 CA Utrecht, The Netherlands
        }

\renewcommand{\maketitlehookd}{%
	\begin{abstract}
		\noindent In astronomy, there is an opportunity to enhance the practice of validating models through statistical techniques, specifically to account for measurement error uncertainties. While models are commonly used to describe observations, there are instances where there is a lack of agreement between the two. This can occur when models are derived from incomplete theories, when a better-fitting model is not available or when measurement uncertainties are not correctly considered. However, with the application of specific tests that assess the consistency between observations and astrophysical models in a model-independent way, it is possible to address this issue. The consistency tests (ConTESTs) developed in this paper use a combination of non-parametric methods and distance measures to obtain a test statistic that evaluates the closeness of the astrophysical model to the observations. To draw conclusions on the consistency hypothesis, a simulation-based methodology is performed. 
        In particular, we built two tests for density models and two for regression models to be used depending on the case at hand and the power of the test needed. We used ConTEST to examine synthetic examples in order to determine the effectiveness of the tests and provide guidance on using them while building a model. We also applied ConTEST to various astronomy cases, identifying which models were consistent and, if not, identifying the probable causes of rejection.
	\end{abstract}
}

\begin{document}

\maketitle 


\section{Introduction} 
\label{sec:intro}

In astrophysics and astronomy, most models depend on a complex combination of physical assumptions and mathematical parameters, making model validation, i.e. testing if the fitted model is a good representation of the observed sample, difficult. In statistics, model validation methods such as the Chi-square test \citep{Snedecor1989}, Kolmogorov-Smirnov test \citep{chakravarti1967}, the Anderson-Darling test \citep{Stephens1974}, etc., are well defined and perform well if the preliminary statistical assumptions hold. However, most well-known statistical tests have preliminary assumptions that are often not satisfied in astronomical settings due to complex multidimensional problems, small sample sizes and observational biases.
Model validation is also strictly dependent on the data used to build the model and the data used to check its goodness of fit. Validation methods based only on the data used to construct the model are often inadequate; in astrophysics, however, the two sets of data often coincide due to the lack of observations for many exotic objects.
In astronomy, it is common to assume that the best-fitting model among those under consideration is an accurate representation of the observations. However, it is worth noting that there may be instances where, if correctly accounting for measurement uncertainties, the chosen model may still be too far from the observations to be statistically consistent.

Model validation is an essential step in the process of building and assessing the performance of a model. There are two main model validation approaches: Bayesian and non-parametric methods. 
In Bayesian model validation, Bayesian statistics principles are employed; in particular, Bayes' theorem is applied to define a parametric description of the problem at hand.  
Bayesian methods are flexible and can handle complex models. Methods like Posterior Predictive assessment of model fitness \citep{Gelman1996,Gelman2013,Lucy2018L} are widely applied by the astronomy community \citep{Kiziltan2013,Feeney2019}. Non-parametric methods, on the contrary, focus on comparing the observed data with the model predictions using a variety of statistical tests and metrics. They do not make any assumptions about the underlying distribution of the data and are often easy to understand and implement. However, goodness-of-fit tests like Chi-square and Kolmogorov-Smirnov, although being used in hundreds of astronomical papers every year, have been proven to have limitations and biases that are often not fully recognized among astronomers \citep{Babu2006}.

In practical applications, if the data distribution is known (Gaussian, Poisson etc.), a parametric bayesian method will be more powerful than a non-parametric one; however, this is often not the case in astronomy. In scenarios where the parametric form is unknown, either due to the complexity of the problem, observational biases, small sample size etc., a wrong assumption on the parametric form will bring flawed results, while a non-parametric method will capture the problem not needing any additional information.

Statisticians often advise using methods from both branches of statistics, Bayesian and Frequentist, to capture a problem entirely. Since a non-parametric set of tests that could deal with both regression and density models, both in low and high dimensional settings, with no initial assumptions whatsoever, was not available in the astronomical literature, we developed ConTEST.
This paper presents a set of new statistical tests to answer the question, "Is the model consistent with the observations?". Less sensitive with respect to goodness-of-fit methods, it can, however, be used in virtually any model validation situation.

ConTEST can efficiently assess, in a model-independent way, the consistency between observations and an astrophysical model in two scenarios: regression and density models. The non-parametric component ensures that no assumption on the parametric form of the model is needed, while the test framework gives the model-independent component. A hypothesis-testing process is set to evaluate the agreement between the model and the observations. It uses a combination of non-parametric methods and distance measures to obtain a test statistic followed by a simulation-based methodology. In the latter, the model is assumed to be the ground truth, and a high number of samples are generated from it; comparing the test statistics of the simulation samples with the original test statistic of the observations allows us to interpret whether the assumption of consistency holds or whether the assumption has been violated.

\begin{figure}[H]
\centering
\includegraphics[scale=0.75]{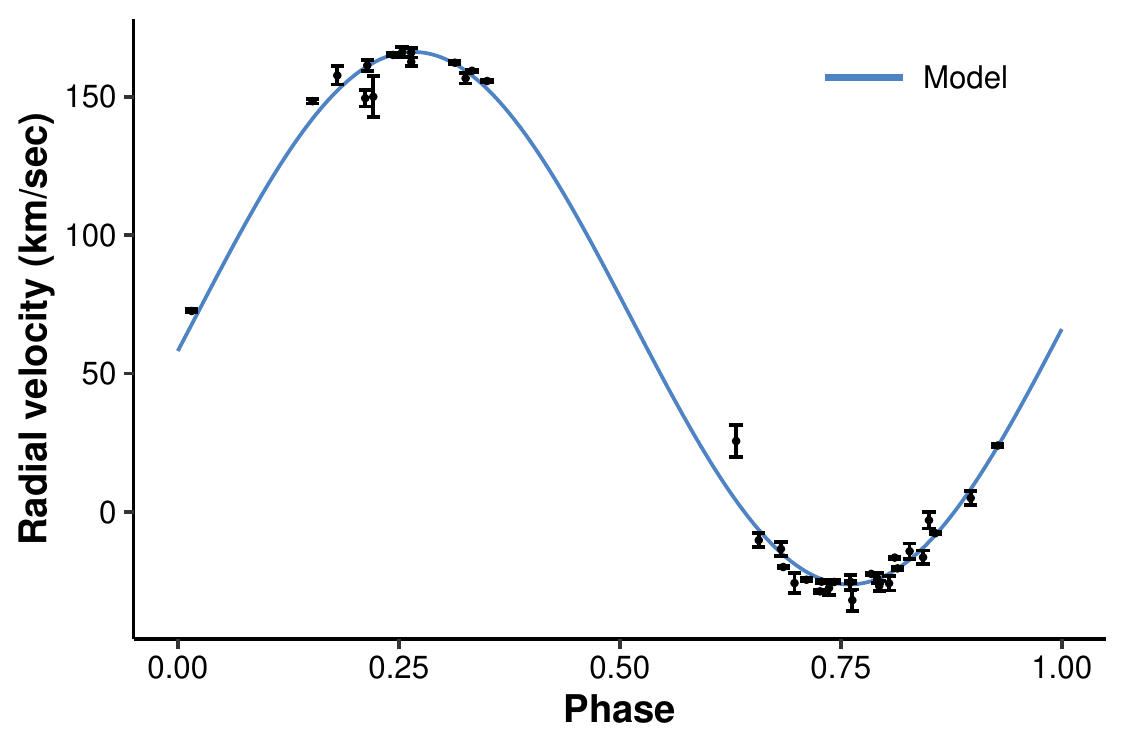}
\caption{The radial velocity vs phase of WD 0326-273 and relative uncertainties (black dots and bars), and the proposed model to evaluate (blue line).}
\label{fig: DoubleWhiteDwarf_onpoints_intro}
\end{figure}

There are two different formulations of ConTEST depending on the type of data and model under consideration. 
In Section \ref{sec: Cons.TestForRegression} we will discuss the structure of ConTEST for regression models. As shown in Fig. \ref{fig: DoubleWhiteDwarf_onpoints_intro}, a typical scenario where this test could be applied is testing the consistency between the modelled radial velocity curve of a White Dwarf and its observations with associated uncertainties.

\begin{figure}[H]
\centering
\includegraphics[scale=0.75]{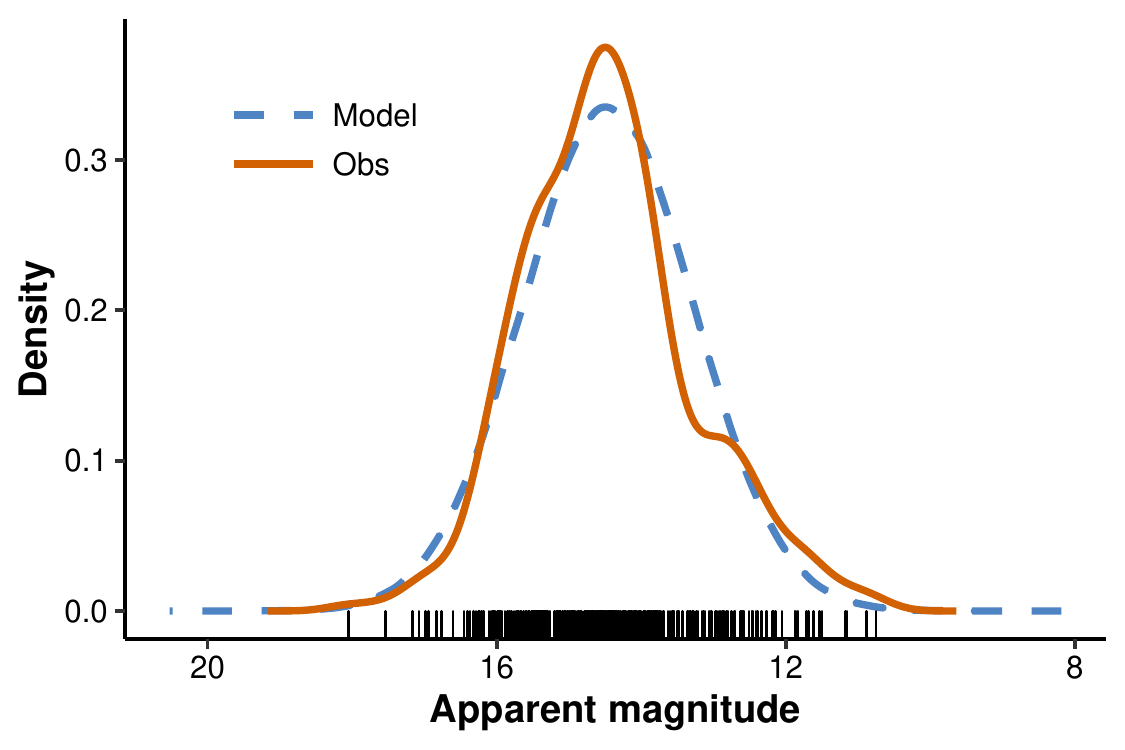}
\caption{The apparent magnitude of 360 GCs in the Andromeda Galaxy (black ticks) and their density (orange line), and the model density (blue dashed line).}
\label{fig: Globular_intro}
\end{figure}

In Section \ref{sec: Cons.TestForDensity}, we will discuss the structure of ConTEST for density models. An example of these scenarios is testing the consistency between the luminosity distribution of globular clusters and a parametric model, as shown in Fig. \ref{fig: Globular_intro}. 

Testing the consistency in a model-independent way makes ConTEST useful both during the construction of a model and to ensure its validity for successive use for inference purposes. The ideal moment to apply ConTEST is when a set of fitted models have been selected, and we want to assess which subset of these models is consistent with the data. 

In the following sections, we will present the consistency tests, explore why a model may be rejected, and investigate the primary component that brought us to this conclusion. The three most common reasons found are a faulty model, an observed biased sample, and wrongly estimated uncertainties. In Section \ref{sec: Example}, we apply ConTEST to multiple real astronomical examples.

\section{Consistency test for regression models}
\label{sec: Cons.TestForRegression}

In astronomy and astrophysics, regression modelling is used both to quantify the relationships in the observed data, like the linear relationship in the Fundamental Plane of elliptical galaxies \citep{Djorgovski1987} or the power-law index of solar flare energies \citep{Hudson1991} and to create physical driven models based on the understanding of a specific phenomenon, giving insights on things like mass, temperature, compositions etc. \citep{feigelson_babu_2012}.
ConTEST can assess the agreement between the observations and the suggested model for both these types of regression.

\subsection{ConTEST for regression}
\label{sec: ConTEST for regression}

We first introduce the main version of the consistency test, ConTEST for regression models. It is applicable for checking the consistency of an observed sample, its associated measurement uncertainties, and a regression model with one dependent variable $Y$ and one or more independent variables $X$. This test is powerful but strict, often rejecting the null hypothesis when outliers, bias, or wrong uncertainty estimation come into play. 

We use the average of the standardised residuals as a distance measure; this test statistic does not assume any parametric form on its distribution, nor any assumption on the data distribution is needed. Furthermore, it is applicable for any homoscedastic or heteroscedastic measurement uncertainties, with the only caveat of being able to generate samples from it.

ConTEST follows these four steps:

\begin{enumerate}
    \item Calculate the distance between the observations, $y_i$ for $i=1,...,n$, and the astrophysical model evaluated at the observations, $\hat{y}_i$, weighted by the observed uncertainties:
    $$ D = \sqrt{ \frac{1}{n} \sum_{i=1}^n \left( \frac{y_{i} - \hat{y}_i}{\sigma_{i}} \right)^{2}},$$
    
    where $\sigma_i$ is the uncertainty of each observation.
    
    \item Simulate $K$ datasets assuming the astrophysical model $\hat{y}$ as ground truth and adding Gaussian noise based on the uncertainties of the observations: 
    $$ y_{ik} = \hat{y}_i + \epsilon_i$$
    where $\epsilon_i \thicksim N(0, \sigma_i) $.

    \item Calculate the distance between the $K$ simulated datasets and the model $\hat{y}$, weighted by the observed uncertainties: 
    $$ D_k = \sqrt{ \frac{1}{n} \sum_{i=1}^n \left( \frac{y_{ik} - \hat{y}_i }{\sigma_{i}} \right)^2 }$$
    
    \item Compare the distribution of the simulations' distances with the test statistic. Reject/not reject the null hypothesis with a significance level $\alpha=0.05$.
\end{enumerate}

\noindent
ConTEST for regression is a two-tailed test. It does not reject the consistency hypothesis when the test statistic falls between the two critical values, that, for a significance level $\alpha = 0.05$, are the $2.5$ and $97.5$ percentiles of the simulations' distances distribution.
Due to the dependence of the test statistic on the observed uncertainties, this test is sensitive to any outlier or wrongly estimated uncertainty. In the steps above, we assume the observed uncertainties to be gaussian, however, any other distribution would work if we can generate samples from it.

To evaluate the effectiveness of the test, we built three synthetic examples for which we know everything: model, observations, and their uncertainties distribution. 
For the true model, we choose an arbitrary function:

\begin{equation}
    f(x) =  \exp(\beta_1 x) \, sin(\beta_2 x)  + m
\label{eq: regression function}
\end{equation} 

\noindent
and for the uncertainties model, $\epsilon(x)$, we simply model the standard deviation at $x$ as a constant fraction of $f(x)$.
The examples simulate three main scenarios: 
\begin{enumerate}
    \item Both the astrophysical model and observations come from the true model and uncertainties, $f(x)$ and $\epsilon(x)$. \\ 
    \item The astrophysical model is biased, i.e. Eqn. \ref{eq: regression function} is fitted without the exponential term. While the observations and relative uncertainties come from the true model and uncertainties, $f(x)$ and $\epsilon(x)$. \\ 
    \item The astrophysical model and the observations come from the true model and uncertainties, $f(x)$ and $\epsilon(x)$. But the estimate of the uncertainties of the observations is underestimated by a factor of $2$. \\
\end{enumerate}

\begin{figure*}[h]
\centering
\includegraphics[scale=.75]{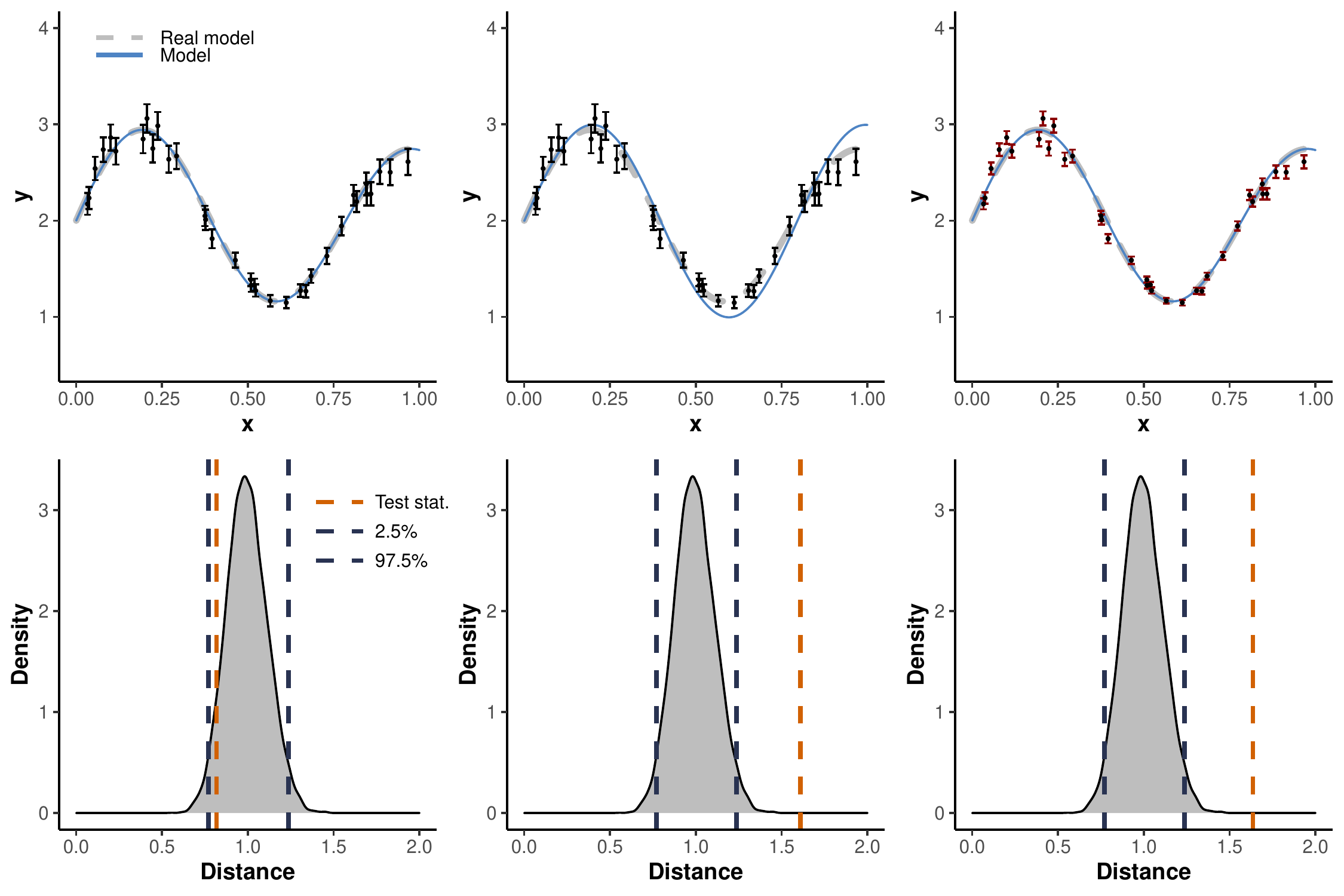}
\caption{Results of ConTEST for regression models; each column represents a scenario. In the first, the model, observations, and uncertainties come from the true model. The test statistic (orange dashed line) lies within the simulations' distances density; the null hypothesis is not rejected.
In the second, the model is biased, while the observations and their uncertainties come from the true model. The test statistic (orange dashed line) lies outside the simulations' distances density; the null hypothesis is rejected.
In the third, the model and the observations come from the true model, but the uncertainties of the observations are underestimated. The test statistic (orange dashed line) lies outside the simulations' distances density; the null hypothesis is rejected.}
\label{fig: points_model}
\end{figure*}

As shown in Fig. \ref{fig: points_model}, ConTEST can correctly not reject the null hypothesis when model, observation and uncertainties are coming from the truth. It can also correctly reject scenarios (2) and (3) where bias and underestimated uncertainties are present. However, ConTEST shows strict behaviour mainly triggered by scenario (3), where the uncertainties are underestimated.
This behaviour is expected even more for real cases where outliers and uncertainty estimation come almost always into play. 

To evaluate the power of the test as a function of the number of observed samples, we repeatedly apply ConTEST to the same three examples for three different sample sizes. Being able to generate observed samples and applying ConTEST a high number of times for each example gives us an estimate of its rejection rate. The results can be found in Table \ref{table: 1}.

\begin{table}[ht]
\caption{Rejection rate of ConTEST for the three examples as a function of the number of observations.}             
\label{table: 1}      
\centering                          
\begin{tabular}{c c c c}        
\hline\hline                 
Example & $n=10$ & $n=100$  & $n=1000$ \\    
\hline                        
   (1) & $5.5\%$ & $4.7\%$ & $5.3\%$  \\
   (2) & $68.3\%$ & $100\%$ & $100\%$ \\
   (3) & $88.1\%$ & $100\%$ & $100\%$ \\
   \hline                                   
\end{tabular}
\end{table}

For scenario (1), the rejection rate is, as expected, close to the significance level $\alpha=0.05$ chosen for the hypothesis testing. This value would converge to $0.05$, increasing the number of repetitions to calculate the rejection rate.
For scenarios (2) and (3), the rejection rate improves as a function of the sample size reaching $100\%$ in both cases with 100 observations.

\subsection{Smoothed ConTEST for regression}
\label{sec: Smoothed ConTEST for regression}

Built specifically to be used in real scenarios where the presence of outliers, bias and problems with the estimated uncertainties come into play, Smoothed ConTEST includes a non-parametric smoothing process to reduce their effects. The non-parametric method used, Local Linear Regression (LLR), is part of the more general family of local polynomial regressors \citep{Nadaraya1964, Muller1998}. 

Local linear regression is a non-parametric method used to estimate the relationship between a dependent variable $Y$ and one or more independent variables $X$. It is particularly useful when the relationship between $X$ and $Y$ is not linear or when the data has high variability. LLR fits a linear regression model to a subset of the data rather than the entire dataset; this subset, also known as a local neighbourhood, is defined by a kernel function. The latter assigns a weight only to the points in the proximity of the prediction location so that distant points do not contribute to the estimate. A specific description of the kernel concept can be found in Section \ref{sec: kde}. 

The main advantage of this non-parametric method is the complete absence of a parametric form, making it an excellent tool for Smoothed ConTEST, where we want to smooth out the observation without any a priori knowledge. The only parameter choice we allow in Smoothed ConTEST is between a fixed or adaptive kernel's bandwidth. An adaptive kernel's bandwidth is especially advised in scenarios where the range of values for the covariates $X$ has gaps; however, for big datasets, the computation time will significantly increase with respect to a fixed one. In both cases, fixed or adaptive, the values of the bandwidth themselves are automatically estimated. 

In Smoothed ConTEST, the LLR not only mitigates the effect of outliers but also provides a confidence interval that is used in the denominator of the new test statistic, effectively removing the dependency of the test on the estimated uncertainties of the observations. However, the measurement uncertainties are still used in the simulation process to perform hypothesis testing.

Smoothed ConTEST follows the same structure of ConTEST presented in Section \ref{sec: ConTEST for regression}. The main difference is that the distance is calculated between the model and the LLR instead of directly from the observations:

\begin{enumerate}

    \item Apply the LLR to the observations $y_i$, for $i=1, ..., n$ and obtain $\hat{y}_{LLR, \; i}$ and its uncertainties $\hat{\sigma}_{LLR, \, i}$.
    \item Calculate the distance between the LLR and the astrophysical model $\hat{y}$:
    $$ D = \sqrt{\frac{1}{n} \sum_{i=1}^n \left( \frac{\hat{y}_{LLR, \; i} - \hat{y}_i}{\hat{\sigma}_{i \, , \, LLR}} \right)^2 }$$
    \item Simulate $K$ dataset assuming the astrophysical model as ground truth and adding Gaussian noise based on the uncertainties of the  observations: 
    $$ y_{ik} = \hat{y}_i + \epsilon_i$$
    where $\epsilon_i \thicksim N(0, \sigma_i) $.

    \item Apply the LLR to each simulated dataset $y_{ik}$ and obtain $\hat{y}_{LLR, \; ik}$ and their uncertainties $\hat{\sigma}_{LLR, \, ik}$.
    
    \item Calculate the distance between each LLR and the astrophysical model $\hat{y}$:
    $$ D_k = \sqrt{\frac{1}{n} \sum_{i=1}^n \left( \frac{\hat{y}_{LLR, \; ik} - \hat{y}_i}{\hat{\sigma}_{ik \, , \, LLR}} \right)^2 }$$
    \item Compare the distribution of the simulations' distances with the test statistic. Reject/not reject the null hypothesis with a significance level $\alpha=0.05$. 
\end{enumerate}

We use the same three examples introduced in Section \ref{sec: ConTEST for regression} to evaluate the power of this new non-parametric consistency test.

\begin{figure*}[h]
\centering
\includegraphics[scale=0.75]{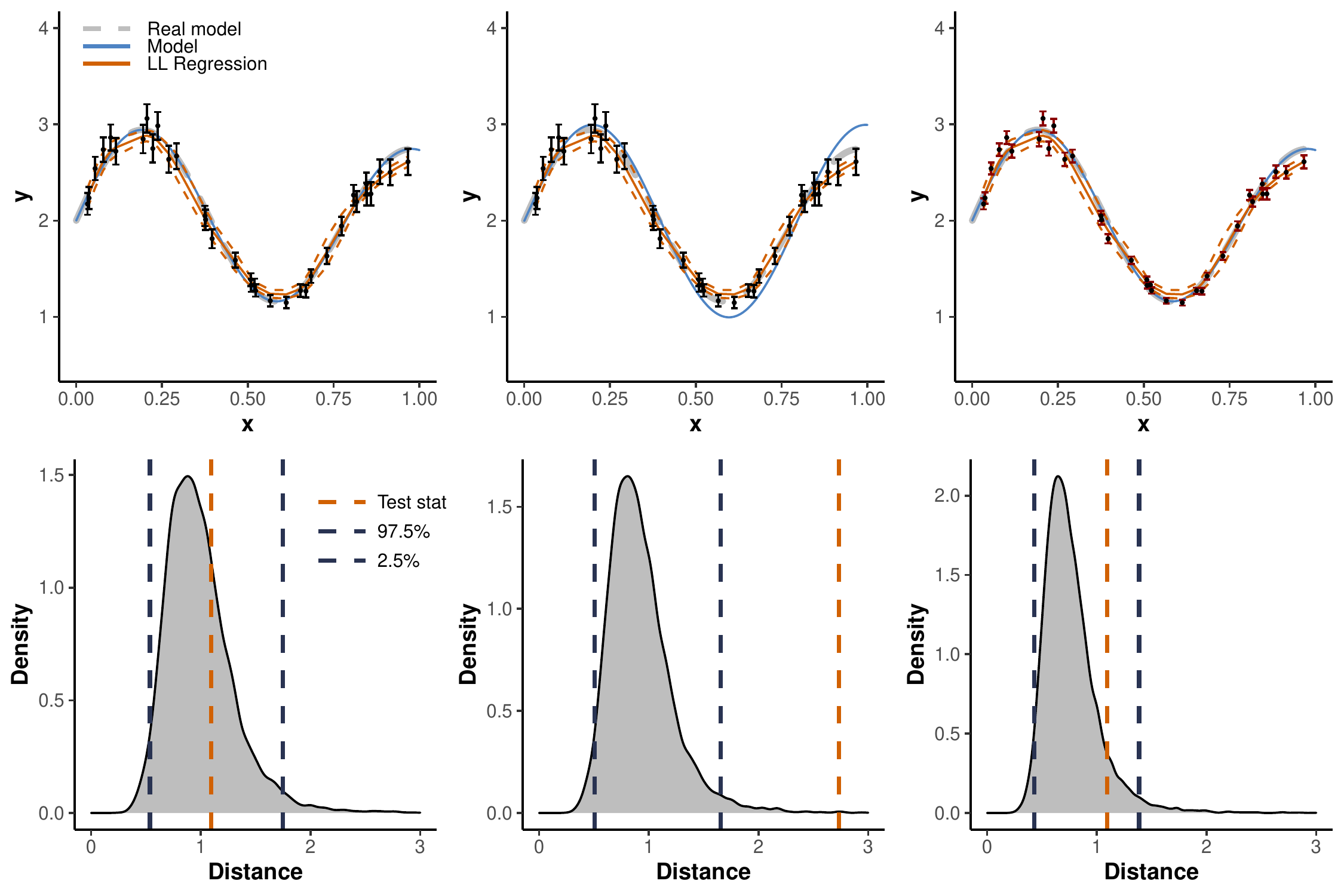}
\caption{Results of Smoothed ConTEST for regression models; each column represents a scenario. In the first, the model, observations, and uncertainties come from the true model. The test statistic (orange dashed line) lies within the simulations' distances density; the null hypothesis is not rejected.
In the second, the model is biased, while the observations and their uncertainties come from the true model. The test statistic (orange dashed line) lies outside the simulations' distances density; the null hypothesis is rejected.
In the third, the model and the observations come from the true model, but the uncertainties of the observations are underestimated. The test statistic (orange dashed line) lies within the simulations' distances density; the null hypothesis is not rejected.}
\label{fig: LLR_model}
\end{figure*}

As shown in Fig. \ref{fig: LLR_model}, Smoothed ConTEST correctly does not reject the consistency hypothesis for scenario (1), where both observations and model are coming from the same ground truth. The biased model of scenario (2) is also correctly rejected, and it is not consistent with the observed data; however, for scenario (3), Smoothed ConTEST is not as strict as ConTEST, and it does not reject the consistency hypothesis when the uncertainties are underestimated by a factor of 2.    
Smoothed ConTEST results in a more relaxed test, being less sensitive to the uncertainties of the observations but a more appealing method for testing the biases of an astrophysical model.

We evaluate the power of the test as a function of the number of observed samples, repeatedly applying Smoothed ConTEST to the three examples for three different sample sizes. The results can be found in Table \ref{table: 2}.

\begin{table}[ht]
\caption{Rejection rate of Smoothed ConTEST for the three examples as a function of the number of observations.}             
\label{table: 2}      
\centering                          
\begin{tabular}{c c c c}        
\hline\hline                 
Example & $n=10$ & $n=100$  & $n=1000$ \\    
\hline                        
   (1) & $5.7\%$ & $4.1\%$ & $5.2\%$  \\
   (2) & $53.7\%$ & $100\%$ & $100\%$ \\
   (3) & $29.8\%$ & $11.2\%$ & $18.1\%$ \\
   \hline                                   
\end{tabular}
\end{table}

For scenario (1), the rejection rate is, as expected, close to the significance level $\alpha=0.05$.
For scenario (2), the rejection rate improves as a function of the sample size, correctly reaching $100\%$ rejection rate with less than 100 observations.
For scenario (3), however, the test cannot reject the null hypothesis independently of the number of observations.

Applying both ConTEST and Smoothed ConTEST on an astrophysical model allows for identifying its weaknesses in terms of biases and under/over-estimation of the observed uncertainties.

\subsection{ConTEST for families of models}
\label{sec: Cons.TestForFamilies}

As briefly introduced in Section \ref{sec: Cons.TestForRegression}, in astronomy and astrophysics, regression modelling is used either to quantify the relationships in the observed data or to create physically driven models based on the understanding of a specific phenomenon. 

ConTEST can be used to assess the consistency of both the specific fit of a model to the given data, as well as the consistency of the overall model family being used.
This method involves re-estimating the model parameters for every simulated dataset during hypothesis testing, allowing for testing a more general model family while accounting for the variations in fit due to measurement uncertainties.

We use ConTEST for regression to explain the framework, however, this variant of the test can also be applied with Smoothed ConTEST:

\begin{enumerate}

    \item Calculate the distance between the observations, $y_i$ for $i=1,...,n$, and the astrophysical model evaluated at the observations, $\hat{y}_i$, weighted by the observed uncertainties:
    $$ D = \sqrt{\frac{1}{n} \sum_{i=1}^n \left( \frac{y_{i} - \hat{y}_i}{\sigma_{i}} \right)^2 },$$
    
    where $\sigma_i$ is the uncertainty of each observation.
    
    \item Simulate $K$ datasets assuming the astrophysical model $\hat{y}$ as ground truth and adding Gaussian noise based on the uncertainties of the observations: 
    $$ y_{ik} = \hat{y}_{i} + \epsilon_i$$
    where $\epsilon_i \thicksim N(0, \sigma_i) $.

    \item Fit the model under consideration to the $K$ simulated datasets, obtaining a set of fitted models $\hat{y}_{ik}$ 
    \item Calculate the distance between the $K$ simulated datasets and their associated model $\hat{y_k}$, weighted by the observed uncertainties: 
    $$ D_k = \sqrt{\frac{1}{n} \sum_{i=1}^n \left( \frac{y_{ik} - \hat{y}_{ik} }{\sigma_{i}} \right)^2 }$$
    
    \item Compare the distribution of the simulations' distances with the test statistic. Reject/not reject the null hypothesis with a significance level $\alpha=0.05$.

\end{enumerate}

\noindent
We show its application in the white dwarf example of Section \ref{sec: RV}.

\section{Consistency test for density models}
\label{sec: Cons.TestForDensity}

Models formulated as densities or as clouds of points from a simulation are often the case in many astrophysical studies; codes like Modules for Experiments in Stellar Astrophysics (MESA, \citealp{Paxton2010}), Compact Object Mergers: Population Astrophysics and Statistics (COMPAS, \citealp{Riley2022}), The Gravitational Wave Universe Toolbox (GWToolbox, \citealp{Shu-Xu2022a}), and many more, often provide as output simulated observations based on a set of input parameters. A way to compare these samples with real observations is not well defined in astronomy, and we hope to create an intuitive framework with ConTEST for densities.

As for the regression case, we want to build a statistical test to evaluate the consistency between the observations and the model in a non-parametric way without specifying any form for either observations or models. To do so, we use a non-parametric method called Kernel Density Estimation (KDE, \citealp{Rosenblatt1956, Parzen1962}), one of the primary data smoothing methods available in statistics. KDE is commonly used to summarize a cloud of points into a continuous distribution allowing us to infer the properties of a population. In astronomical settings, it has been used, for instance, in the study of star clusters \citep{Seleznev2016}, the study of binary black holes mergers \citep{Shu-Xu2022b}, and often as a data representation method being preferable to histograms \citep[e.g.][]{Weglarczyk2018}.

In Section \ref{sec: kde}, we briefly explain the idea behind the KDE, and in Section \ref{sec: density_simplified} and \ref{sec: framework_density}, we present two tests: ConTEST for outliers and ConTEST for densities.

\subsection{Kernel density estimation}
\label{sec: kde}

Here we present the kernel density estimation and its use in our consistency test. 
For $(X_1, X_2, \ldots, X_n)$ independent and identically distributed variables, KDE is defined as

\begin{equation}
\hat{f}(x,h)={\frac {1}{n}}\sum _{i=1}^{n}K_{h}(x-X_{i})={\frac {1}{nh}}\sum _{i=1}^{n}K{\Big (}{\frac {x-X_{i}}{h}}{\Big )},
\label{eq: KDE}
\end{equation}

\noindent
where $K$ is the kernel and $h > 0$ is the smoothing parameter called bandwidth. Due to its convenient mathematical properties, we chose a Gaussian kernel, $K(x) = \Phi(x)$, for ConTEST; however, a range of kernel functions are available such as triangular, biweight, triweight, Epanechnikov, etc. 
The choice of the kernel does not affect the density estimate if the number of observations is high enough; however, the value of the bandwidth $h$ does. The most common methods to estimate the bandwidth are rules-of-thumb like Scott's rule \citep{Scott1987, Scott1992} $H = \left(n \right)^{-2/(m+4)} \Sigma$ and Silverman's rule \citep{Silverman1988} $H = \left( \frac{4} {n(m + 2)} \right) ^{2/(m+4)} \Sigma$, where $m$ is the number of variables, n is the sample size and $\Sigma$ is the empirical covariance matrix; for our application, we use Scott's rule being on average good for most scenarios.

KDE is easily extendable to the multivariate case making it an excellent tool for our non-parametric tests. To estimate densities $f$ in $R^p$, it simply performs an average of multivariate kernels centred at the data points. One disadvantage of KDE is that it is not accurate when estimating a density near the finite endpoints of the support, e.g. near $0$ for a strictly positive variable. A set of solutions for boundary-corrected KDE are available and have been explored in \citealt{Jones1993} and \citealt{Karunamuni2005}; however, an easy solution to the problem is transforming the data such that the hard boundaries are removed, e.g. applying a logarithmic transformation to both model and observations.

\subsection{ConTEST for outliers}
\label{sec: density_simplified}

In astronomy, a natural approach often found in the literature is calculating the likelihood of an observed sample with respect to the density model under consideration and using it as a test statistic. The test introduced below is based on the same principle; however, we argue that, due to the formulation of its test statistic, this is not a consistency test but an outliers detection method.
ConTEST for outliers follows these steps:

\begin{enumerate}
    \item If the model is given as a simulated cloud of points: apply KDE to the model sample and obtain its density $g$.
    \item Calculate the loglikelihood of the observations $y_i$, for $i=1,...,n$, with respect to the density $g$: 
    \begin{equation}
        D = - \frac{1}{n} \sum_{i=1}^{n} log( \, g(y_i) \,)
    \end{equation}
    \item Simulate $K$ datasets $y_{ik}$ of size $n$ from the model's density $g$.
    \item Calculate the loglikelihood for each simulated dataset with respect to the density $g$:
        \begin{equation}
        D_{k} = - \frac{1}{n} \sum_{i=1}^{n} log( \, g(y_{i \, k}) \,)
        \end{equation}
    \item Compare the distribution of simulations' distances with the test statistic. Reject/not reject the null hypothesis with a significance level $\alpha=0.05$. 
\end{enumerate}

\noindent
The resulting test is simple and can be used with a low amount of observations. However, this test is not a consistency test but an outlier detection method. The reason for this is that there exist distributions $f \neq g$ such that if $Y\sim g$ and $X\sim f$, we have that $E\left[-log(g(Y))\right] \approx E \left[-log(g(X)) \right]$. This means that even if we have a considerable amount of data, we would not be able to reject $g$ when in fact, $f$ is the true distribution. Of course, it is rather unlikely that the true distribution has this property with respect to $g$. Furthermore, an outlier would have a very low likelihood and, therefore, would be easily detected with this method.


We evaluate the effectiveness of this test with synthetic examples. For the true model, we chose an arbitrary bivariate Gaussian distribution $\mathcal{N}(\boldsymbol{\mu}_{true},\,\boldsymbol{\Sigma}_{true})\,$

\noindent
with mean
$$ {\boldsymbol{ \mu}_{true}} = \left[ 
\begin{array}{c}  5 \\ 5  
\end{array} \right] $$

\noindent
and covariance matrix

$$ \boldsymbol{\Sigma}_{true} = \left[
\begin{array}{cc}
1.5 & 0.8 \\
0.8 & 2.5 \\
\end{array}
\right]. $$

The examples simulate three main scenarios:

\begin{enumerate}
    \item Both the astrophysical model and observations come from the true model.\\ 
    \item The astrophysical model is biased, $${\boldsymbol{\mu}_{mod}} = \left[ \begin{array}{c} 4 \\ 4  \\ \end{array} \right].$$
    While the observations come from the true model.\\ 
    \item The astrophysical model has an overestimated covariance matrix, 
    $$\boldsymbol{\Sigma}_{mod} = \left[
    \begin{array}{cc}
    2.5 & 0.8 \\
    0.8 & 3.5 \\
    \end{array}
    \right].$$

    While the observations come from the true model. \\ 
    
\end{enumerate}

\begin{figure*}[h]
\centering
\includegraphics[scale=0.75]{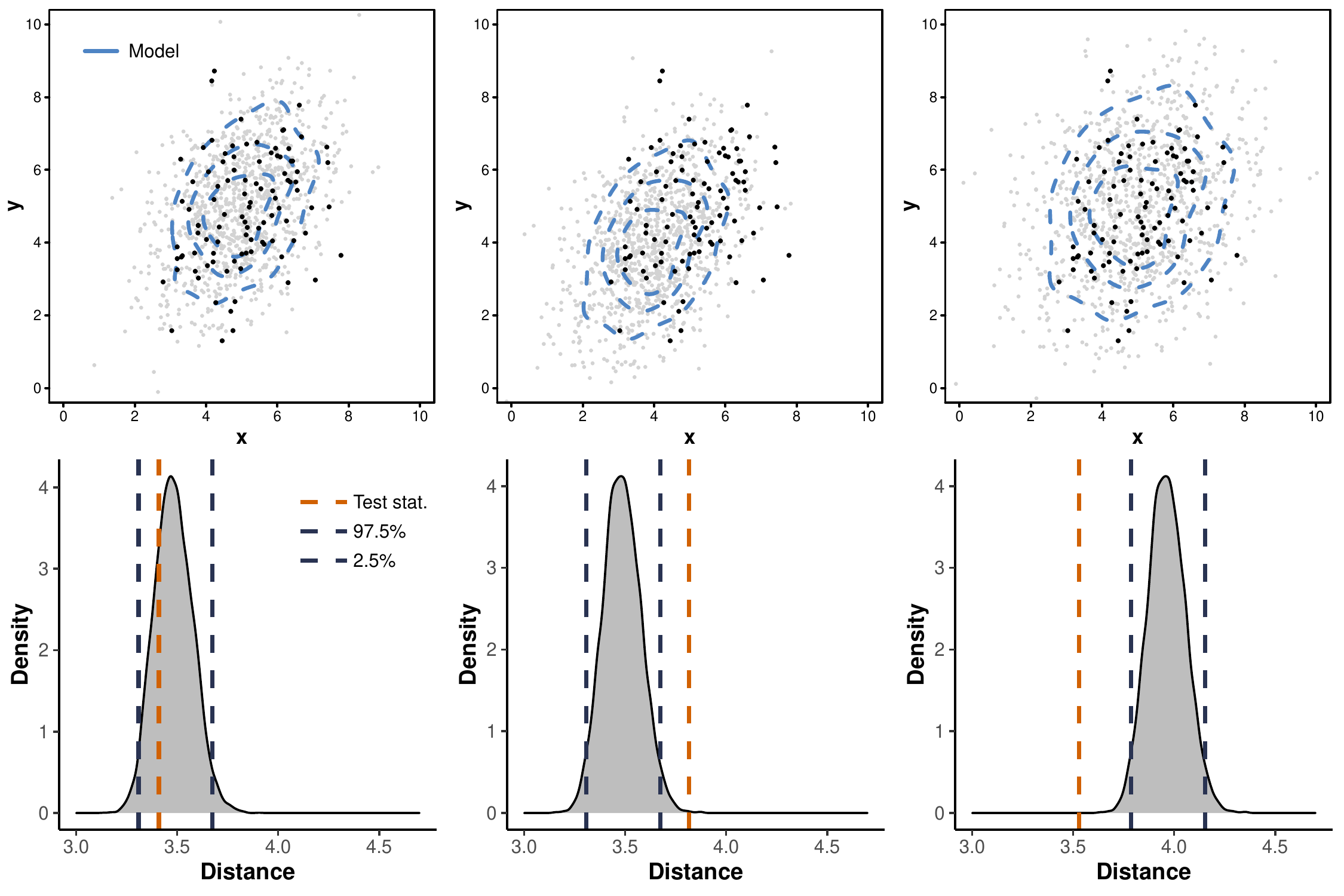}
\caption{Result of ConTEST for outliers; each column represents a scenario. In the first, the model and observations come from the true model. The test statistic (orange dashed line) lies within the simulations' distances density; the null hypothesis is not rejected.
In the second, the model is biased, while the observations come from the true model. The test statistic (orange dashed line) lies outside the simulations' distances density; the null hypothesis is rejected.
In the third, the astrophysical model has an overestimated covariance matrix, while the observations come from the true model. The test statistic (orange dashed line) lies outside the simulations' distances density; the null hypothesis is rejected.}
\label{fig: density_model_outliers}
\end{figure*}

As can be seen in Fig. \ref{fig: density_model_outliers}, the test correctly does not reject scenario (1), while it does reject scenarios (2) and (3).

Although we call it an outlier detection method, this test can also reject in the opposite scenario where an unlikely high number of observations are lying in a high-density region of the model.


As for the regression tests, we repeatedly apply the test to the three examples for three different sample sizes to evaluate the power of ConTEST for outliers as a function of the number of observations. Repeating the test a thousand times for each scenario gives Table \ref{table: 3} rejection rates.

\begin{table}[ht]
\caption{Rejection rate of ConTEST for outliers for the three examples as a function of the number of observations.}             
\label{table: 3}      
\centering                          
\begin{tabular}{c c c c}        
\hline\hline                 
Example & $n=10$ & $n=100$  & $n=1000$ \\    
\hline                        
   (1) & $5.3\%$ & $5.4\%$ & $4.9\%$ \\
   (2) & $18.4\%$ & $90.7\%$ & $100\%$ \\
   (3) & $25.1\%$ & $99.4\%$ & $100\%$ \\
   \hline                                   
\end{tabular}
\end{table}

For scenario (1), the rejection rate is, as expected, approximately equal to the significance level $\alpha=0.05$.
For scenarios (2) and (3), the rejection rate improves as a function of the sample size and both reach more than $90\%$ rejection rate with 100 observations.

\subsection{ConTEST for densities}
\label{sec: framework_density}

Here we introduce the framework to test the consistency between observations and astrophysical models with ConTEST for densities. The test uses only non-parametric methods and a distance measure to reject/not reject the null hypothesis. 
ConTEST follows these steps:

\begin{enumerate}
    \item If the model is given as a simulated cloud of points: Apply KDE to the model sample and obtain its density $g$.
    \item Apply KDE to the observed sample and obtain its density $f$.
    \item Calculate the distance between $f$ and $g$ with:
    \begin{equation}
        D = \int_{\mathbb{R}} \lvert{f(y)-g(y)}\lvert \,dy
    \end{equation}
    \item Simulate K dataset of size n from the model's density $g$.
    \item Apply KDE to each of the K datasets and obtain their densities $f_k$.
    \item Calculate the distance for each simulated dataset:
        \begin{equation}
        D_{k} = \int_{\mathbb{R}} \lvert{f_k(y)-g(y)}\lvert \,dy
    \end{equation}
    \item Compare the distribution of simulations' distances with the test statistic. Reject/not reject the null hypothesis with a significance level $\alpha=0.05$. 
\end{enumerate}

The test statistic is approximated with an MCMC method simulating a high number of observations from $g$ and then calculating the distance with $ D= \frac{1}{N} \sum_{i=1}^N  \lvert\frac{f(x_i)}{g(x_i) } -1 \lvert  $. Due to the formulation of its test statistic, ConTEST for density is a one-tailed test; it rejects the consistency hypothesis only when the test statistic is greater than the critical value.

\noindent
We use the same three synthetic examples introduced in Section \ref{sec: density_simplified} to test the power of ConTEST for density models.

\begin{figure*}[h]
\centering
\includegraphics[scale=0.7]{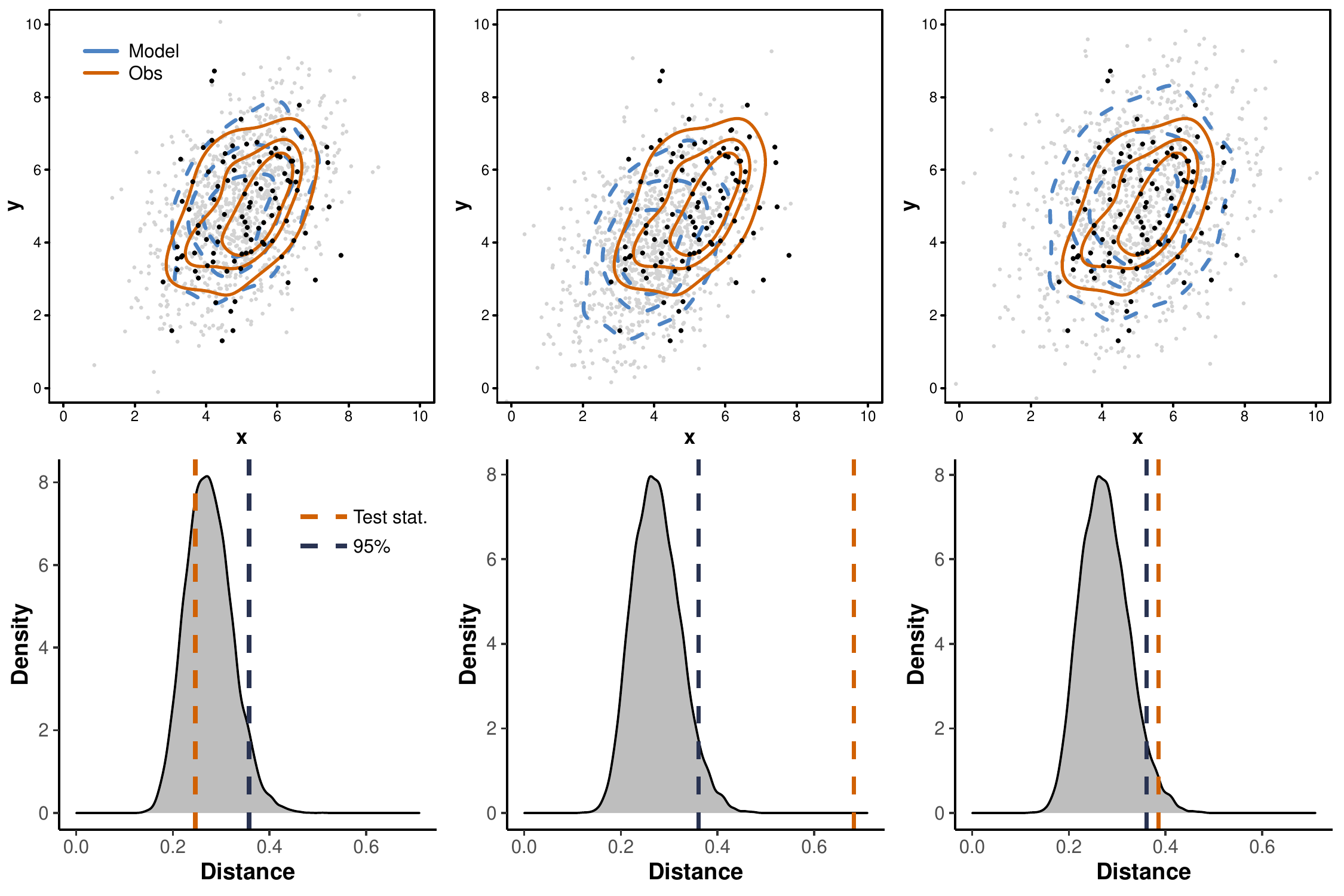}
\caption{Result of ConTEST for densities; each column represents a scenario. In the first, the model and observations come from the true model. The test statistic (orange dashed line) lies within the simulations' distances density; the null hypothesis is not rejected.
In the second, the model is biased, while the observations come from the true model. The test statistic (orange dashed line) lies outside the simulations' distances density; the null hypothesis is rejected.
In the third, the astrophysical model has an overestimated covariance matrix, while the observations come from the true model. The test statistic (orange dashed line) lies outside the simulations' distances density, and the null hypothesis is rejected.}
\label{fig: density_model}
\end{figure*}

As shown in Fig. \ref{fig: density_model}, for scenario (1), where both model and observations come from the truth, ConTEST for densities does not reject the consistency hypothesis. For scenario (2) correctly rejects the consistency hypothesis being the model biased. And in scenario (3), it can assess that the model has an overestimated covariance matrix making it not consistent with the observed sample. 

Also, for ConTEST for densities, we evaluate the power of the test dependency on the number of observations; the results can be found in Table \ref{table: 4}.

\begin{table}[ht]
\caption{Rejection rate of ConTEST for densities for the three examples as a function of the number of observations.}             
\label{table: 4}      
\centering                          
\begin{tabular}{c c c c}        
\hline\hline                 
Example & $n=10$ & $n=100$  & $n=1000$ \\    
\hline                        
   (1) & $4.9\%$ & $4.6\%$ & $5.2\%$ \\
   (2) & $40.8\%$ & $100\%$ & $100\%$ \\
   (3) & $12.0\%$ & $31.1\%$ & $100\%$  \\
   \hline                                   
\end{tabular}
\end{table}

For scenario (1), the rejection rate is, as expected, approximately equal to the significance level $\alpha=0.05$.
For scenario (2), the rejection rate improves as a function of the sample size and reaches $100\%$ rejection rate with 100 observations. For scenario (3), the test shows low sensitivity to overestimating the covariance and only rejects the null hypothesis $31.1\%$ of the time. However, with a bigger sample size, it reaches a $100\%$ rejection rate.

\section{Examples}
\label{sec: Example}

We now present a series of examples to show the applicability of ConTEST to real scenarios in astronomy and, at the same time, evaluate some well-known astrophysical models.

\subsection{Cosmic Microwave background}

The first example concerns the Cosmic Microwave Background (CMB) radiation from which essential properties of the Universe as a whole and its evolutionary history can be determined.  The many sky surveys of the CMB, such as the Wilkinson Microwave Anisotropy Probe (WMAP \citealp{Bennett2013, Hinshaw2013}, Planck \citep{Adam2014}), the South Pole Telescope (SPT, \citealp{Story2013}) and the Atacama Cosmology Telescope (ACT, \citealp{Sievers2013}), although agreeing on a standard $\Lambda CDM$ model with six cosmological parameters, derive different parameter estimates. This is a clear case where testing the consistency of the models with the observations is necessary.
The relevant data is the power spectrum of the temperature fluctuations measured over the sky, which shows several peaks. We apply both ConTEST methods for regression to test the consistency of the Planck 2018 temperature power spectrum data \citep{Aghanim2020} with the best-fitting model made available to the public from the Planck collaboration.

\begin{figure*}[h]
\centering
\includegraphics[scale=0.7]{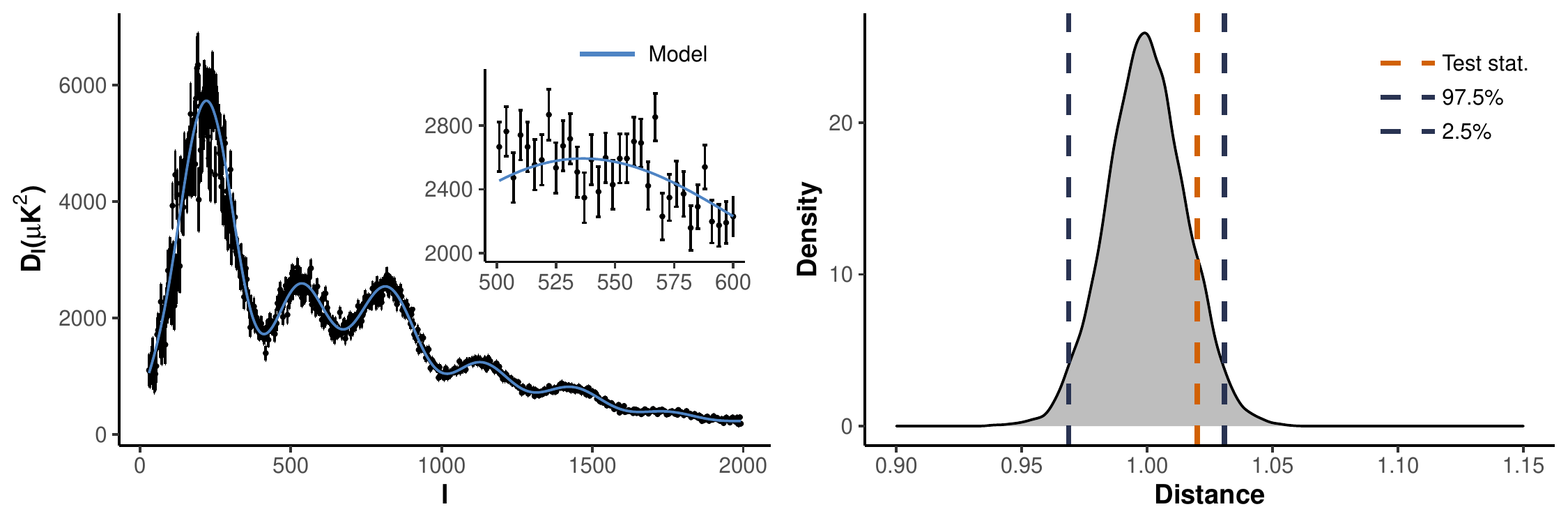}
\caption{Results of ConTEST. On the left are the Planck 2018 temperature power spectrum data, relative uncertainties (black dots and bars), and the publicly available best-fitting model (blue line). On the right, the test statistics (orange dashed line) lie within the simulations' distances density; the consistency hypothesis is not rejected.}
\label{fig: CMB2018TT_onpoints}
\end{figure*}

\begin{figure*}[h]
\centering
\includegraphics[scale=0.7]{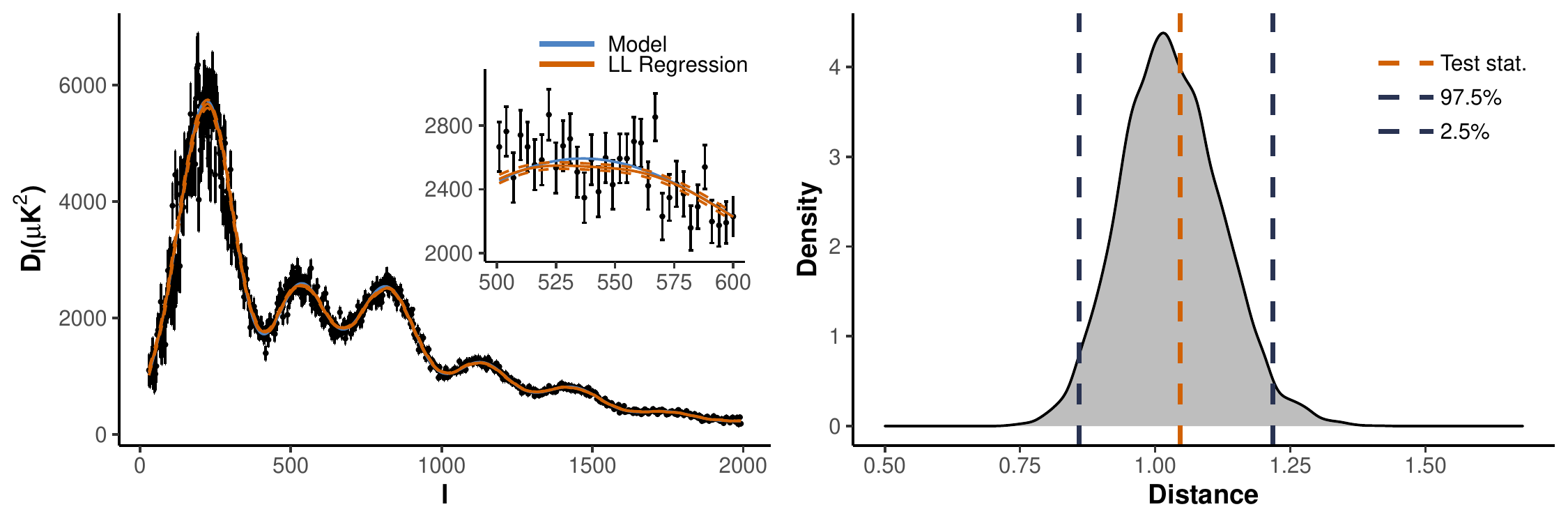}
\caption{Results of Smoothed ConTEST. On the left, the Planck 2018 temperature power spectrum data and relative uncertainties (black dots and bars), the publicly available best-fitting model (blue line) and the LLR model (orange line). On the right, the test statistics (orange dashed line) lie within the simulations' distances density; the consistency hypothesis is not rejected.}
\label{fig: CMB2018TT}
\end{figure*}

In Fig. \ref{fig: CMB2018TT_onpoints}, we show the data, and the best-fitting model on the left and on the right ConTEST result shows the consistency of the model with the data. In Fig.\ref{fig: CMB2018TT}, we show the LLR approach's results that are very close to the best-fit model and also lead to the conclusion that the model is consistent with the data. As expected, Planck's $\Lambda CDM$ model is consistent with the observations and does not suffer from either bias or under/over-estimation of uncertainties.

\subsection{Spectral analysis}

A similar type of data set occurs in spectroscopy, where the intensity of light as a function of wavelength provides essential information about the temperature and chemical composition of the observed object and shifts in velocity due, e.g. to doppler motion, can be detected. It is common practice to develop different synthetic spectra based on an object's temperature, composition and other properties and compare them to the observed data to select the best-fit model. This is a clear example where a consistency test helps validate the theoretical model. 

Here investigate a typical spectra analysis and fitting from \cite{Martocchia2021}, which determine C and N abundances in stars of star clusters in the Magellanic Clouds. We test its consistency with Smoothed ConTEST.

\begin{figure*}[h]
\centering
\includegraphics[scale=0.7]{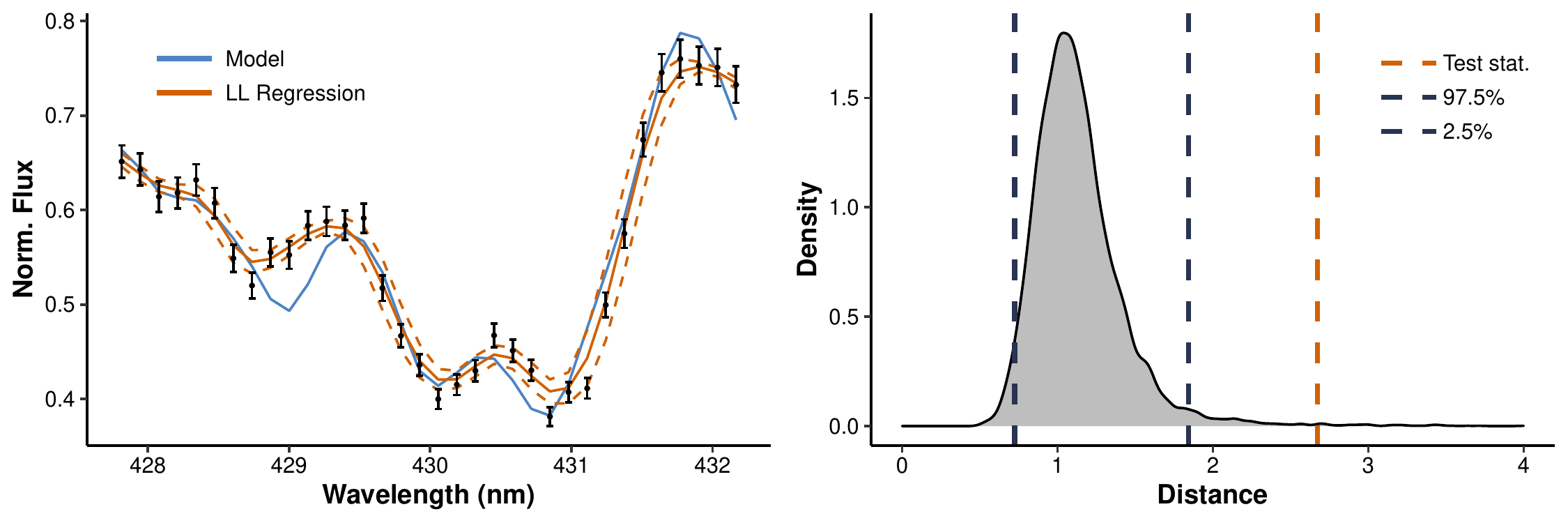}
\caption{Result of Smoothed ConTEST. On the left, the normalized flux spectrum and relative uncertainties (black dots and bars), the proposed model to evaluate (blue line) and the LLR model (orange line). On the right, the test statistics (orange dashed line) lie outside the simulations' distances density, and the consistency hypothesis is rejected.}
\label{fig: Spectra}
\end{figure*}

In Fig. \ref{fig: Spectra}, we show a part of the normalized spectrum around 430 nm, where absorption from CH is present. The best-fit model is shown with the data and the LLR result. Although the best fit matches the data for a significant part of the spectral range, ConTEST rejects the best fit. In this case, as often in theoretical modelling, the model does not fully capture the complexity of the data. Astronomers can use ConTEST to test the consistency and comment on the outcome, making the limitations of the theoretical model explicit.

\subsection{Binary white dwarf radial velocity}
\label{sec: RV}

In the following example, we test the consistency between the follow-up observations of a double white dwarf binary detected in the ESO SN Ia Progenitor SurveY (SPY) and its orbital solution model from \cite{Nelemans2005}. In particular, we apply both ConTEST methods for regression to the binary white dwarf (WD) $0326-273$, consisting of a close double white dwarf with a possible outer M star third companion in a very wide orbit. 

This example differs from the ones introduced before; here, the model comes from simple Newtonian laws and thus is very unlikely to be wrong or incomplete, while the observations, often combined data from different telescopes, and with uncertainties estimated a posteriori, are the main suspects of possible rejection.

\begin{figure*}[h]
\centering
\includegraphics[scale=0.7]{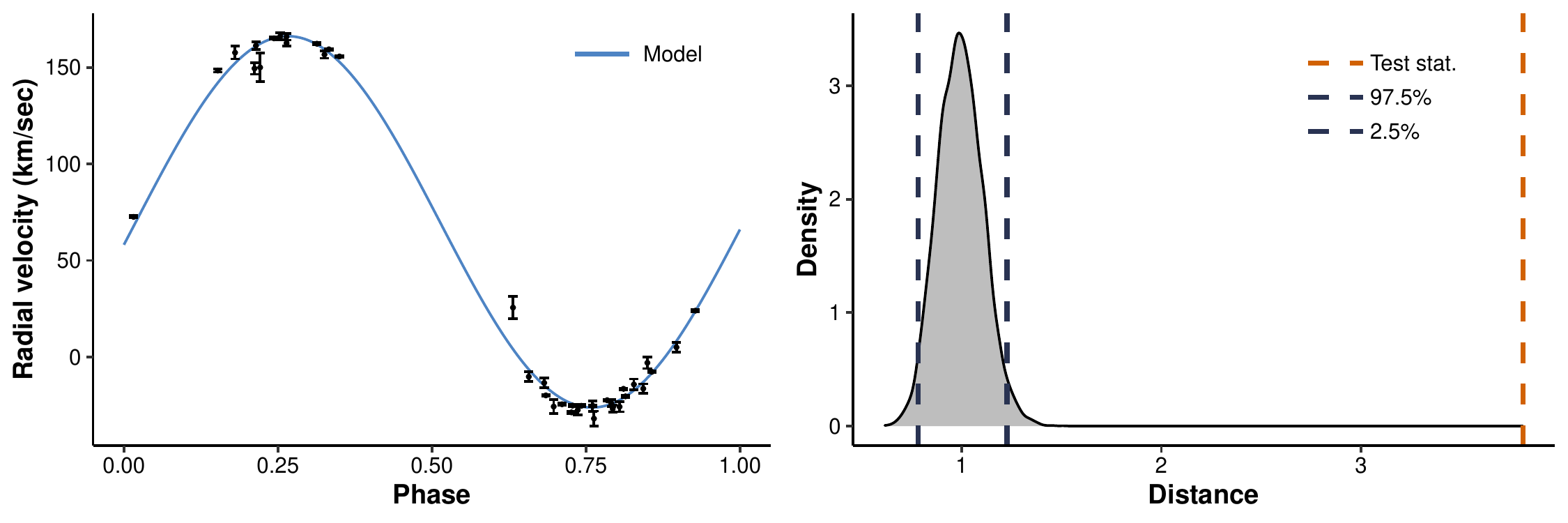}
\caption{Result of ConTEST. On the left is the radial velocity vs phase of WD$0326-273$, relative uncertainties (black dots and bars), and the proposed model to evaluate (blue line). On the right, the test statistics (orange dashed line) lie outside the simulations' distances density, and the consistency hypothesis is not rejected.}
\label{fig: DoubleWhiteDwarf_onpoints}
\end{figure*}

\begin{figure*}[h]
\centering
\includegraphics[scale=0.7]{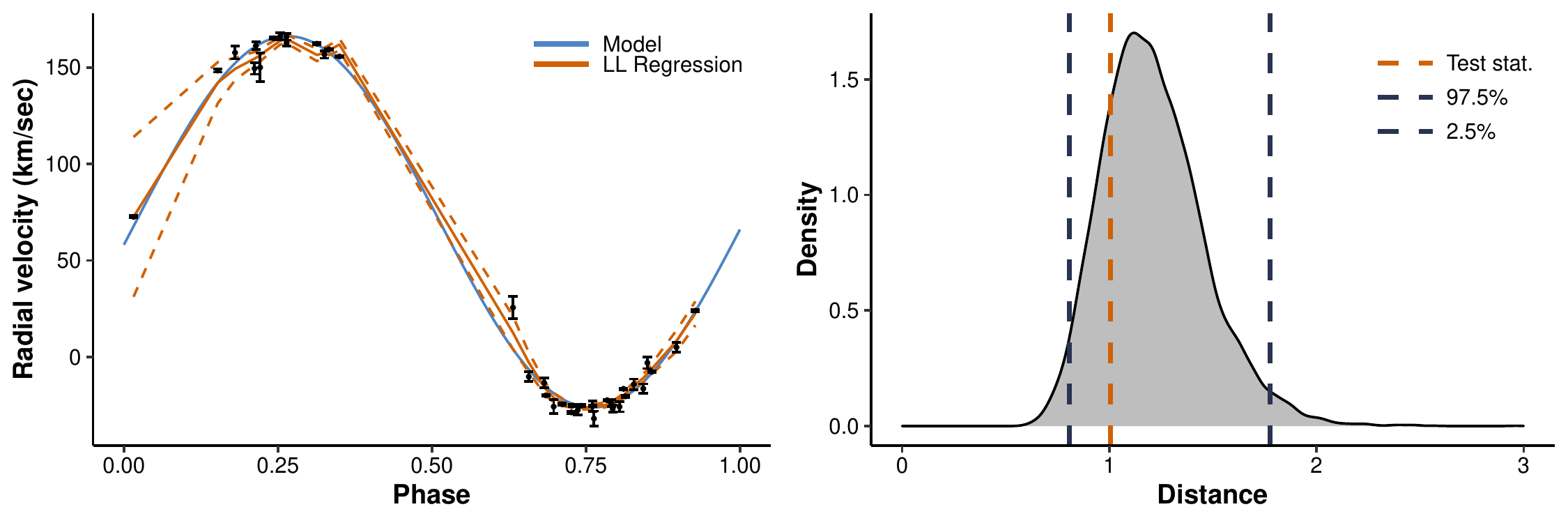}
\caption{Result of Smooted ConTEST. On the left, the radial velocity vs phase of WD$0326-273$ and relative uncertainties (black dots and bars), the proposed model to evaluate (blue line) and the LLR model (orange line). On the right, the test statistics (orange dashed line) lie inside the simulations' distances density; the consistency hypothesis is not rejected.}
\label{fig: DoubleWhiteDwarf}
\end{figure*}

First, in  Fig. \ref{fig: DoubleWhiteDwarf_onpoints}, we show the data and the best-fit model from the paper. ConTEST for regression rejects the consistency hypothesis. This is likely due to the observations' tiny uncertainties and possible systematic errors and offsets, not the model itself. Indeed, testing the model again with Smoothed ConTEST, which is less sensitive to the effect of the observed uncertainties, the consistency hypothesis is not rejected, further confirming that the rejection cause is indeed the uncertainties. This result can be found in Fig \ref{fig: DoubleWhiteDwarf}.

\begin{figure*}[h]
\centering
\includegraphics[scale=0.7]{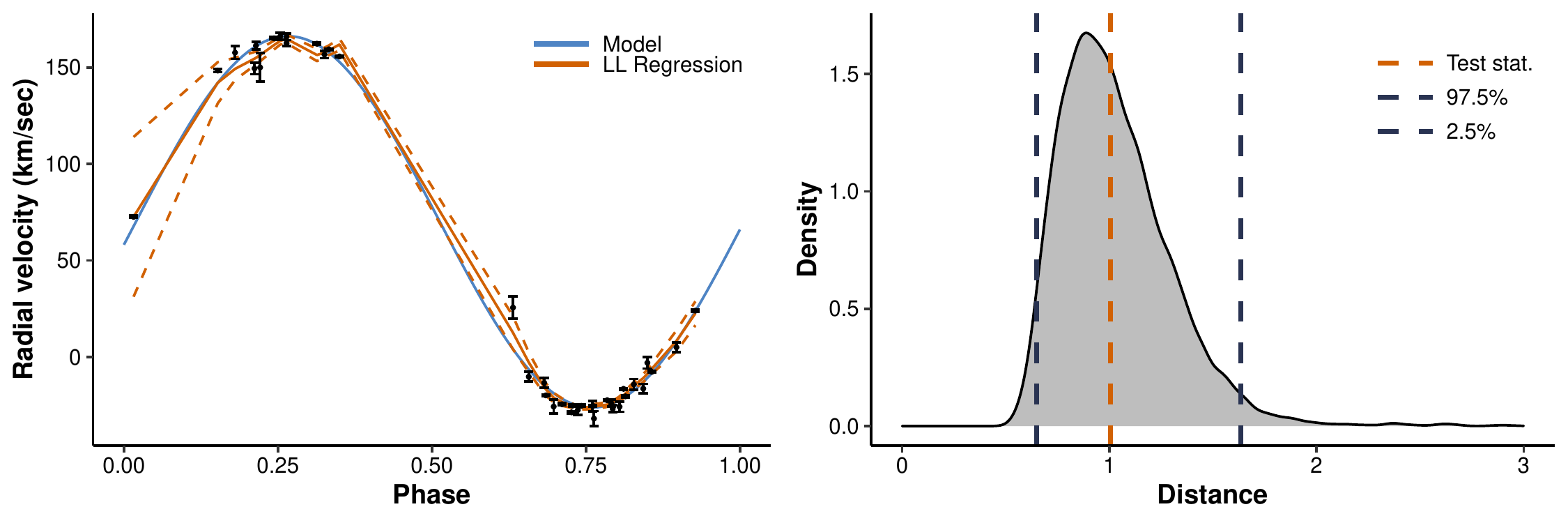}
\caption{Result of ConTEST for family of model. On the left, the radial velocity vs phase of WD $0326-273$ and relative uncertainties (black dots and bars), the proposed model to evaluate (blue line) and the LLR model (orange line). On the right, the test statistics (orange dashed line) lie inside the simulations' distances density; the consistency hypothesis is not rejected for the family of the model.}
\label{fig: DoubleWhiteDwarf_family}
\end{figure*}

For this example, we also apply Smoothed ConTEST for the family of models coming from Newtonian laws. Since the model is estimated directly from the observations, we can re-estimate a new model for every simulated dataset. The result for this test is shown in Fig. \ref{fig: DoubleWhiteDwarf_family}, and it strongly hints that the family of models is indeed correct, not rejecting the consistency hypothesis between the model and the observations of the binary WD $0326-273$.

We also used this specific example to test the adaptive bandwidth; being the support unevenly sampled, the LLR with an adaptive bandwidth can produce a smoother estimate of the regression function.

\subsection{Globular cluster luminosity}

We now apply ConTEST on density models. As a first example, we look at the Globular Cluster (GC) luminosity function. GCs are large collections of stars gravitationally bound in a compact configuration, fundamentally distinct from the field population of stars in the same galaxy. GCs can offer clues regarding different aspects of their hosting galaxies, such as the galactic star formation history, the cannibalism of smaller merging galaxies, and the galactic structure \citep{Fall1988, Ashman1998, West2004}. 
 
The distribution of GC luminosities, known as the globular cluster luminosity function (GCLF), can provide essential insights for inferring cosmological distances \citep{Racine1968, Hanes1977}. 
Early analyses of the Milky Way's GCLFs and the Andromeda Galaxy (M 31) showed that a Gaussian distribution was an excellent analytical fit to the observed distribution of luminosities \citep{Racine1979, Harris1991}. Further analyses indicated many more possible analytical formulas that better fit the observed luminosities; however, the problem is beyond the purpose of this example and will not be discussed here.
We test the initial Gaussian fit proposed for the GCLF of 360 GCs in the Andromeda Galaxy made available in \citealt{Nantais2006}.

\begin{figure*}[h]
\centering
\includegraphics[scale=0.7]{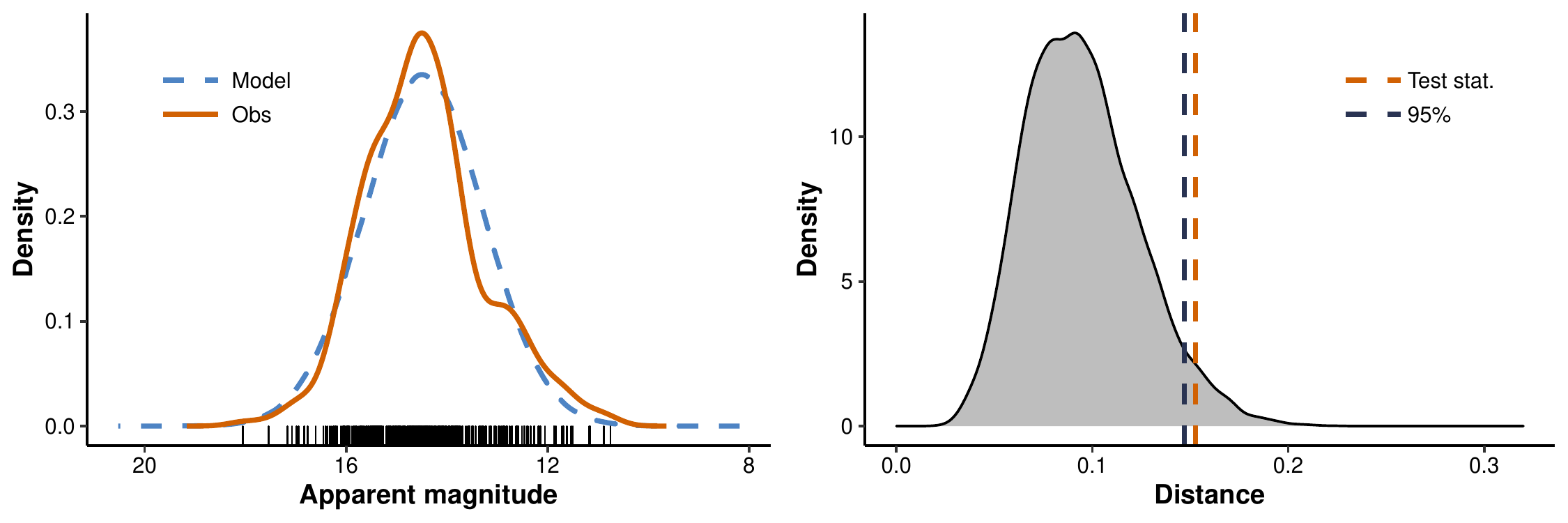}
\caption{Result of ConTEST for densities. On the left, the apparent magnitude of 360 GCs in the Andromeda Galaxy  (black ticks) and their density (orange line), and the model density (blue dashed line). On the right, the test statistic (orange dashed line) lies beyond the critical value; the null hypothesis is rejected.}
\label{fig: Glob_cluster}
\end{figure*}

In Fig. \ref{fig: Glob_cluster}, we show the distribution of GC luminosities and the best fit Gaussian model. As shown in the figure, ConTEST for densities rejects the Gaussian fit. This is an excellent example of a model that comes from the best-fitting set of parameters estimated from the observations but is still too incomplete to be considered consistent.

\subsection{Double neutron star population synthesis}

Finally, we give an example of a two-dimensional distribution. Since the discovery of the Hulse-Taylor double neutron star (DNS) binary, the number of known DNS in the Milky Way has increased, opening to more accurate population analyses \citep{Tauris2017}. In particular, the properties of the systems are often analysed in the period -- eccentricity plane. According to \citet{Andrews2019}, the current sample hints at the presence of three distinct sub-populations based on their orbital characteristics: (i) short-period, low-eccentricity binaries; (ii) wide binaries; and (iii) short-period, high-eccentricity binaries.

We have produced several synthetic populations based on different assumptions for the direct progenitors of DNS and specific properties of the supernova in which the second neutron star is formed (Fontein et al. in preparation). We use ConTEST to test the consistency of three alternative models for the first sub-population. Because the number of observations is low, we only apply ConTEST for outliers.

\begin{figure*}[h]
\centering
\includegraphics[scale=0.7]{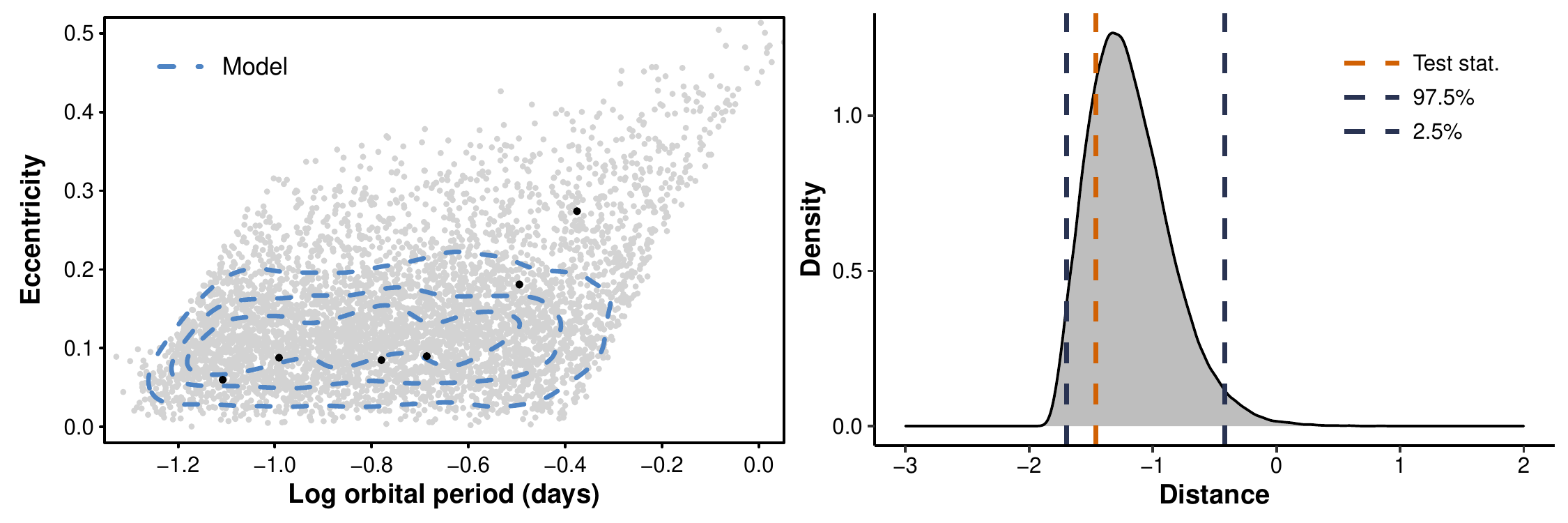}
\caption{Result of ConTEST for outliers. On the left, the eccentricity vs orbital period of the observed DNS (black dots), the eccentricity vs orbital period of the modelled DNS (grey dots), and the model density (blue dashed contours). On the right, the test statistics (orange dashed line) lie within the simulations' distances density; the null hypothesis is not rejected.}
\label{fig: DNS_n1_simple}
\end{figure*}

\begin{figure*}[h]
\centering
\includegraphics[scale=0.7]{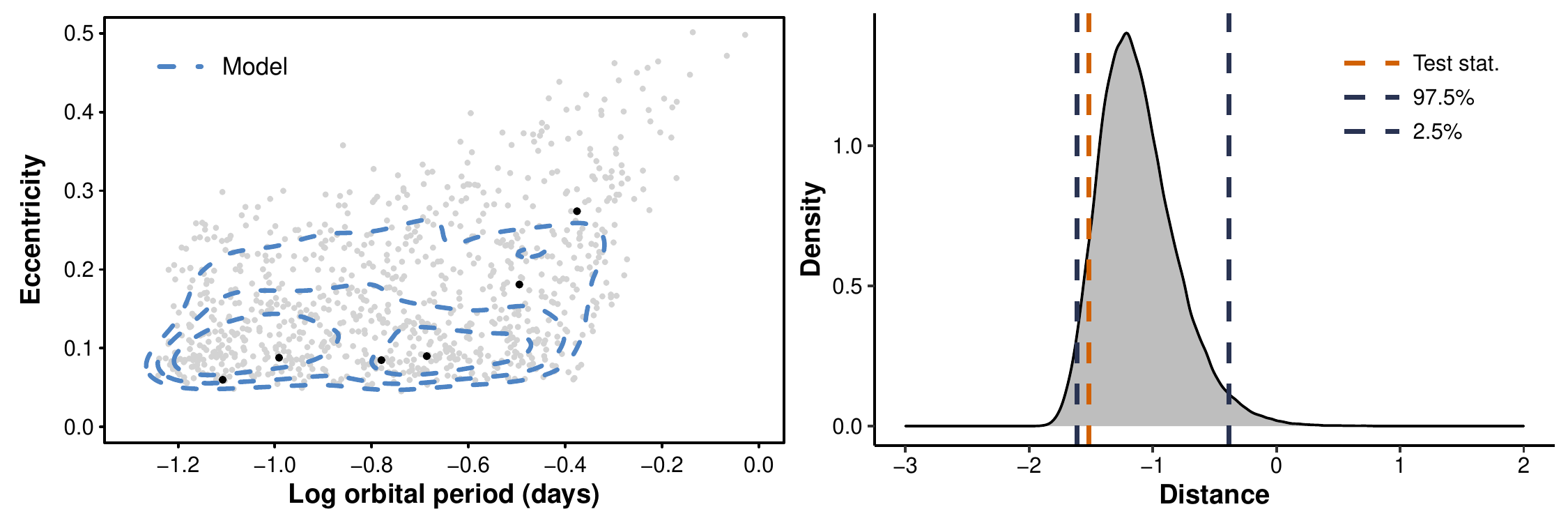}
\caption{Result of ConTEST for outliers. On the left, the eccentricity vs orbital period of the observed DNS (black dots), the eccentricity vs orbital period of the modelled DNS (grey dots), and the model density (blue dashed contours). On the right, the test statistics (orange dashed line) lie within the simulations' distances density; the null hypothesis is not rejected.}
\label{fig: DNS_n3_simple}
\end{figure*}

\begin{figure*}[h]
\centering
\includegraphics[scale=0.7]{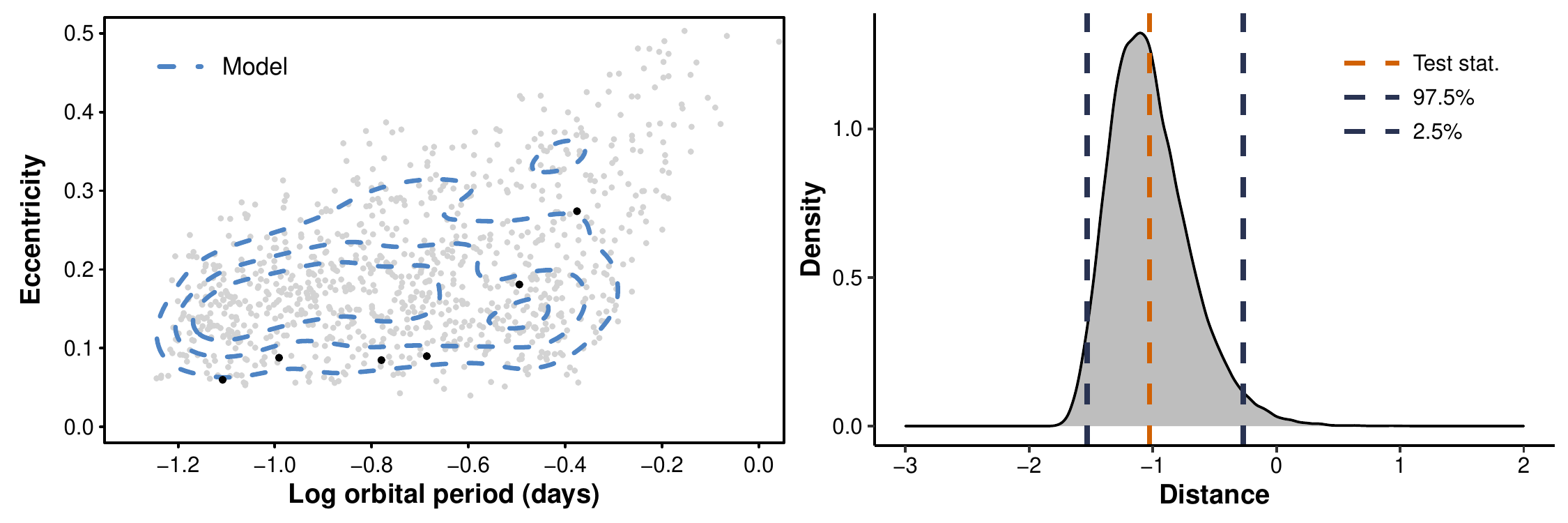}
\caption{Result of ConTEST for outliers. On the left, the eccentricity vs orbital period of the observed DNS (black dots), the eccentricity vs orbital period of the modelled DNS (grey dots), and the model density (blue dashed contours). On the right, the test statistics (orange dashed line) lie within the simulations' distances density; the null hypothesis is not rejected.}
\label{fig: DNS_n2_simple}
\end{figure*}

As it can be seen in Fig \ref{fig: DNS_n1_simple}, \ref{fig: DNS_n3_simple}, and \ref{fig: DNS_n2_simple}, ConTEST for outliers does not reject any of the three possible sub-populations models.


\section{Discussion}

Astronomical data often comes with an estimate of the so-called measurement uncertainties. However, the use of estimated uncertainties in model validation is not the same for regression and density models. In this paper, for regression models, the estimated uncertainties of the observations are a fundamental part of the two tests allowing the creation of simulated samples from the model under consideration. 
The same cannot be done in validation methods for density models. However, two possible ways to include uncertainties in our density consistency test exist: correct the observed density accounting for the measurement uncertainties or modify the model to account for the uncertainties.
The first method is a more accurate approach, although its practical application is still under discussion in the statistical literature with the concept of Deconvolution Kernel Density Estimation \citep{Stefanski1990, Delaigle2008}.
The second approach is certainly easier to apply once an estimate of the measurement error distribution is obtained from theory or observations. It can be used in combination with ConTEST, simply convolving the model sample with the estimated error, getting a broader model that includes the information of measurement uncertainties. 
Either solution makes a better comparison between models and observations and is always advised if measurement uncertainties are significant.

We also want to point out that if a set of possible models is being tested, the consistency tests proposed in this paper can create a confidence region for the parameter space under consideration. Testing a grid of the model's parameters and looking at the subset that is not rejected will automatically create a multidimensional ellipse that can be used as confidence intervals for the parameters.

\section{Conclusions}

In this paper, we propose multiple frameworks for assessing the consistency between observations and astrophysical models in a non-parametric and model-independent way. We developed four tests, two for regression and two for density models. 

For regression models, we tested both methods on three synthetic examples exploring possible problems that can affect an astrophysical model and the observations under consideration. The first method, ConTEST for regression, is powerful but often strict, rejecting the null hypothesis in real cases, especially when under/overestimation of the uncertainties is present; the second method, Smoothed ConTEST for regression, instead, is developed specifically for real applications and it is less sensitive to the uncertainties of the observations, enabling it to concentrate on testing the presence of biases between model and observations. Applying both methods allows evaluation of the three most common problems that affect regression models in astronomy: bias in the model, outliers in the observations, and incoherence of the uncertainties of the observations.

For density models, the first method, ConTEST for outliers, allows for assessing the presence of outliers in the observations with respect to the model under consideration. The second test, ConTEST for densities, is an excellent tool for evaluating the consistency between density models, either parametric or simulated clouds of points, and observations. Applying both tests allows first to assess the presence of outliers, eventually remove them, and then draw conclusions on the consistency of the model.

All four tests were applied extensively on real cases in Section \ref{sec: Example}, and comments on the results were given as guidelines for possible users of this set of non-parametric tests.

In the future, we will concentrate on expanding the use of observations uncertainties while testing models. We will specifically explore the use of uncertainties in testing the consistency of density models and the possibility of including the model's uncertainties in the framework when available.

\section*{Acknowledgements}

F. S. and G. N. acknowledge support from the Dutch Science Foundation NWO.

\section*{Data Availability}

All four ConTEST methods are developed in the statistical software R and available to the public in Python at \url{https://github.com/FiorenSt/ConTEST} \citep{Stoppa2022}.


\bibliography{Bibliography} 
\bibliographystyle{abbrvnat}


\end{document}